%% file: auppap.tex
\documentstyle[sprocl,epsfig]{article}

\def\Journal#1#2#3#4{{#1} {\bf #2}, #3 (#4)}

\def\NPB{{\em Nucl.\ Phys.\ }B}
\def\PLB{{\em Phys.\ Lett.\ }B}

\def\PRD{{\em Phys.\ Rev.\ }D}

\newcommand{\DD}{{\rm D}}

\newcommand{\half}{{\textstyle{\frac{1}{2}}}}

\newcommand{\fin}{\hspace{0.6cm}}
\newcommand{\imag}{{\rm i\hspace{0.13ex}}}

\newcommand{\bgeq}{\begin{equation}}
\newcommand{\bgeqa}{\begin{eqnarray}}
\newcommand{\edeq}{\end{equation}}
\newcommand{\edeqa}{\end{eqnarray}}

\newcommand{\ainv}{a^{-1}}

\newcommand{\chib}{\overline{\chi}_b}
\newcommand{\lqcd}{\Lambda_{\rm QCD}}

\title{Heavy quarks in non-relativistic lattice QCD}
\author{JOACHIM HEIN}
\address{Department of Physics \& Astronomy, University of Glasgow,\\ 
Glasgow G12 8QQ, Scotland, UK}

\begin{document}
\vspace*{-2cm}
\begin{flushleft}
\textbf{\textsf{GUTPA/98/10/1}}\\
\textbf{\textsf{hep-ph/9810314}}\\[2ex]
\end{flushleft}
\maketitle
\abstracts{I give an overview of phenomenological heavy quark results
obtained in NRQCD on the lattice. In particular I discuss the
bottomonium and the $b$-light hadron spectrum.
I also review recent results on the decay constants 
$f_B$ and $f_{B_s}$.}
\section{Introduction}
\subsection{Heavy quarks on the lattice}
Lattice QCD provides an approach to calculate the properties of
hadronic bound states of strongly interacting matter from first
principles.  However when interested in hadronic states involving
heavy $c$ and $b$ quarks standard lattice methods would lead to large
discretisation errors. This is caused by the Compton wave length of the
heavy quark being non-negligible against the lattice spacing. In
present simulations the latter is on the order of a few inverse GeV.

However the Compton wave length of the heavy quarks is an irrelevant scale
for the dynamics of heavy hadrons, see e.g.~the lecture
note~\cite{cschlad}.  
One possibility to simulate heavy quarks
is the  expansion of the heavy quark action around its
non-relativistic limit, which is known as non-relativistic QCD
(NRQCD)~\cite{nrqcd1,nrqcd2}. Another cure is the heavy clover
approach~\cite{fermilab} which is applied by several groups
presently. Within this talk we will review some of the phenomenological
results obtained in NRQCD.

\subsection{Non-relativistic QCD on the lattice} 
In NRQCD the Hamiltonian of the heavy quark is expanded around its
non-relativistic limit
\bgeq\label{hexpan}
H = H_0 + \delta H\,,\fin H_0 = -\frac{\DD^2}{2M_Q}\,.
\edeq
With $H_0$ we denote the leading kinetic term
and $\delta H$ stands for the relativistic and spin dependent 
corrections. In case of a
quarkonium system up to order ${\cal O}(M_Qv_Q^4)$ these corrections
read
\bgeqa\label{hcorrect}
\delta H &=& -{ c_1}\,\frac{(\DD^2)^2}{8M_Q^3}
+ { c_2}\, \frac{\imag g}{8M_Q^2}(\vec \DD\cdot \vec E - \vec E\cdot \vec
\DD)\nonumber \\
&&-{ c_3}\, \frac{g}{8M_Q^2}\vec 
\sigma(\vec \DD \times \vec E - \vec E\times\vec \DD)
-{ c_4}\, \frac{g}{2M_Q}{\vec \sigma\cdot \vec B}\,.
\edeqa
The coefficients $c_i$ have to be determined, such that $H$ in
eqn.~(\ref{hexpan}) matches the Hamiltonian of QCD. So far this has been
investigated in lattice perturbation theory \cite{Morningstar}.
Since NRQCD is non-renormalisable, the $c_i$ will
diverge, if the lattice spacing $a$ is sent to zero. Therefore the
lattice spacing has to be kept finite and improvement terms have
to be added to the action, such that residual discretisation effects
become negligible against other sources of error
\bgeq \label{hdisc}
\delta H_{\rm disc}=
{ c_5}\,a^2 \frac{\sum_i\DD^4_i}{24M_Q}
-{c_6}\,a\frac{(\DD^2)^2}{16nM_Q^2}\,.
\edeq
The $n$ is a stability parameter used in 
equation~(\ref{evolution}). Present calculation use
tadpole improved~\cite{tadpole} tree-level values for all $c_i$'s.
The Hamiltonian (\ref{hexpan}) leads to a differential equation being
first order in time. Therefore the heavy 
quark propagator $G_t$ can be obtained
from an evolution equation
\bgeqa\label{evolution}
G_{t+1}&=& \left(1-a \half \delta H-a\half \delta H_{\rm disc}\right)
\left(1-a {\textstyle\frac{1}{2n}}H_0\right)^n U^+_4\nonumber\\
&&\cdot\left(1-a {\textstyle\frac{1}{2n}}H_0\right)^n
\left(1-a \half \delta H-a\half \delta H_{\rm disc}\right)G_{t}\,.
\edeqa
Thich is numerically quite inexpensive when
compared to the matrix inversions associated with a Dirac type Hamiltonian.

\section{Heavy quarkonia}

\subsection{Bottomonium splittings}\label{upsplit}
The radial and orbital splittings in the bottomonium ($\Upsilon$) and 
charmonium ($J/\Psi$) system are $\approx 500$~MeV, which
approximately equals  the average kinetic energy of the
quarks~\cite{nrqcd1}. 
One obtains $v_Q^2/c^2 = 0.1$ for the $\Upsilon$ and 0.3 for the
$J/\Psi$, which
justifies the non-relativistic approach. 
In the case of heavy quarkonia the
Hamiltonian has to be expanded in powers of 
$v_Q$ instead of $1/M_Q$~\cite{nrqcd2}.

\begin{figure}
\centerline{\protect\epsfig{file=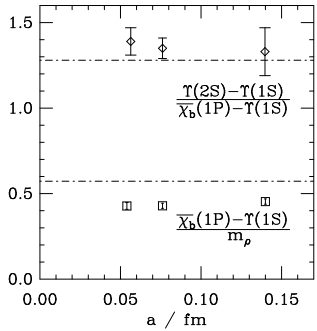,height=4.3cm,
bbllx=62pt,bblly=32pt,bburx=502pt,bbury=502pt}}
\protect\caption{\label{upfig}Scaling of spin independent
splittings~\protect\cite{scaling} in the quenched approximation. 
The points with error bars represent the lattice result, 
the horizontal line the experimental values from PDG \protect\cite{pdg}}
\end{figure}

In figure~\ref{upfig} we show the dependence of spin independent
bottomonium splittings on the lattice spacing $a$.
This calculation has been performed in the quenched approximation,
which neglects the vacuum polarisation by quark-antiquark pairs.
Within the given error bars the lattice result is indeed independent of the
lattice spacing. This shows, that within the achieved accuracy
continuum results can be obtained at finite $a$. In case of
the ratio of the $\overline{\chi}_b(1P)-\Upsilon(1S)$ 
energy splitting~\footnote{$\chib$ denotes the spin average $(\chi_{b0}+
3\chi_{b1}$ + 5$\chi_{b2})/9$} 
to the $\rho$-meson mass~\cite{ukqcdmrho} 
we observe a clear missmatch to the experimental 
result.
This is expected to be caused by the different
running of the strong coupling $\alpha_s$ in the quenched
theory and the {\em real world}, where vacuum polarisation is present.
This expectation is supported by the 
the ratio of the splittings
$[\Upsilon(2S)-\Upsilon(1S)]/[\overline{\chi}_b(1P)-\Upsilon(1S)]$ being
closer to the experimental outcome. In this case both quantities probe
similar momentum scales.

For spin dependent quantities the equation (\ref{evolution})
is not improved to the same level with respect to higher
order relativistic and discretisation corrections as for spin independent
ones. Therefore at this level one observes some scaling violation in
the bottomonium fine and hyperfine
structure~\cite{scaling}. Also the agreement of the $\chi_b$ fine
structure with experiment is not as good as for the radial and orbital
splittings. Improvement needs the inclusion of the higher order
terms~\cite{trottierspin,mankespin} and a better determination of the
$c_i$'s in the Hamiltonian~\cite{c4pap}.

\subsection{Partly unquenching}
\begin{figure}
\centerline{\protect\epsfig{file=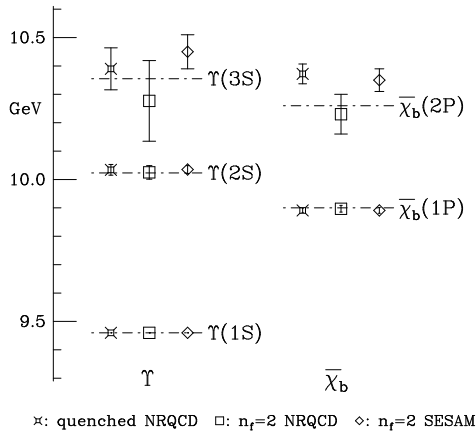,height=5.7cm,
bbllx=40pt,bblly=75pt,bburx=530pt,bbury=520pt}}
\protect\caption{\label{quenfig} Effect of dynamical fermions on the spin
independent $\Upsilon$-spectrum.
The points with error bars represent the lattice
results,
the horizontal line the experimental values from PDG
\protect\cite{pdg}. 
The $a$ has been determined such that the average of the
$\overline{\chi}_b(1P)-\Upsilon(1S)$ and
$\Upsilon(2S)-\Upsilon(1S)$ splitting fits to experiment. One obtains
$\ainv \approx 2.4$~GeV in all cases.
The quenched
results are taken from the NRQCD
collaboration~\protect\cite{scaling}. Squares use 2 flavours of
staggered sea quarks with $am_{\rm sea}=0.01$~\protect\cite{cdyn},
whereas diamonds use Wilson sea quarks extrapolated to 
$m_{sea}=m_s/3$~\protect\cite{achim}.}
\end{figure}

The effect of quenching in quarkonium spectroscopy has been
investigated by two groups~\cite{alphapap,sesam}. Due to algorithmic
reasons, both studies use two flavours of dynamical quarks. In
figure~\ref{quenfig} we compare the outcome for the spin independent
spectrum to the quenched approximation.
Within error bars no significant sea quark effect can be shown.

\subsection{Heavy hybrid mesons}
Hybrid mesons denote colour neutral states consisting of gluonic
excitations as well as a quark-antiquark pair, see e.g. the review by
F.~Close \cite{close}.
Of special interest are so called {\em exotic} states with quantum
numbers unavailable to $q\bar q$-mesons, which are hoped to be the first
identified experimentally.

So far exploratory studies have been undertaken to investigate the
$b\bar b g$ spectrum \cite{manke_hyb,sara_hyb} in NRQCD. 
These use $H_0$ in eqn.~(\ref{evolution}) only. In this approximation
one obtains two sets of degenerate states, which we denote by their
cubic group representation $T_1^{+-}$ and $T_1^{-+}$.
States below the $B\bar
B^{**}$-threshold are expected to be stable. Using the $B_J^*(5732)$
for this threshold one obtains 11.01~GeV. With the $a$-value from
figure~\ref{quenfig} one gets for the masses~\cite{manke_hyb}
\bgeq
M(T_1^{+-})=11.09(10){\rm \ GeV}\,,\fin
M(T_1^{-+})=11.15(5){\rm \ GeV}\,.
\edeq
The result~\cite{sara_hyb} $M(T_1^{+-})=10.82(25)$~GeV is in
agreement with the above. With this size of an error the question of
stability of these states can not be answered.

\section{Heavy light systems}

\subsection{Heavy light spectrum}
The physics of heavy light mesons is quite different from
the physics of heavy quarkonia. For the former in the limit
$M_Q\to\infty$ the heavy quark $Q$ decouples from the dynamics and
becomes a static colour source. The properties of such a state would
be determined entirely by the light quarks and the glue. 
In NRQCD this leads to different power counting 
rules~\cite{arifarules}. The corrections in eqn.~(\ref{hcorrect}) have
to be ordered in powers of $(\lqcd/M_Q)$.

\begin{figure}[tp]
{
{\protect\epsfig{file=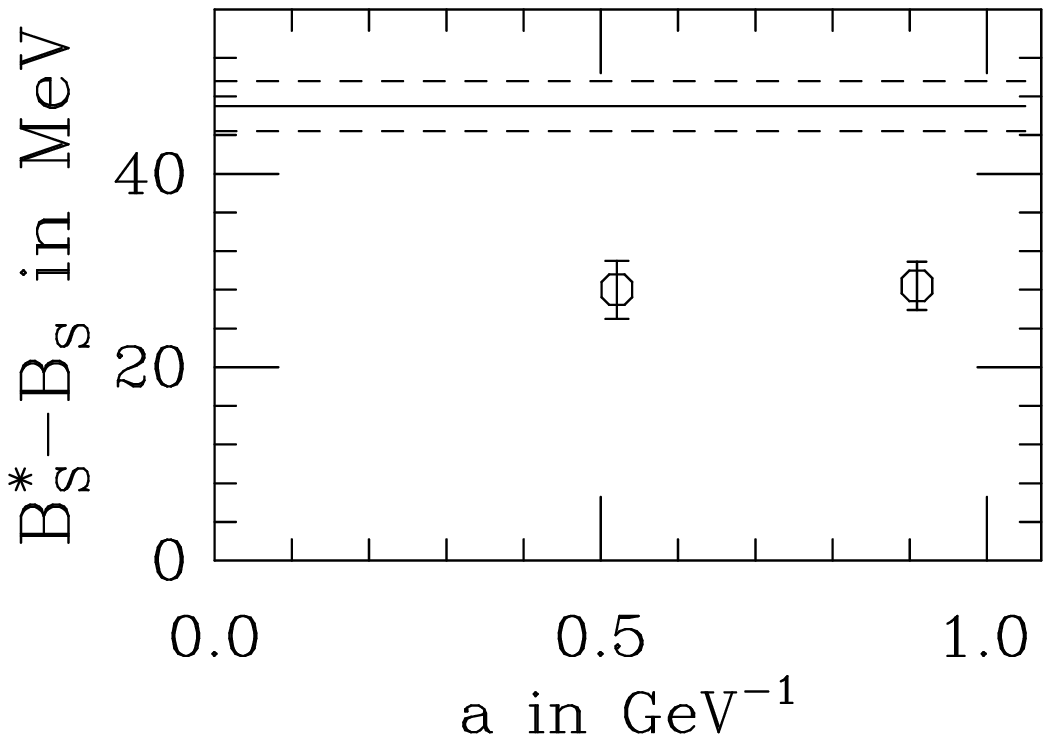,
width=3.7cm,bbllx=35pt,bburx=350pt,bblly=55pt,bbury=265pt}}
{\protect\epsfig{file=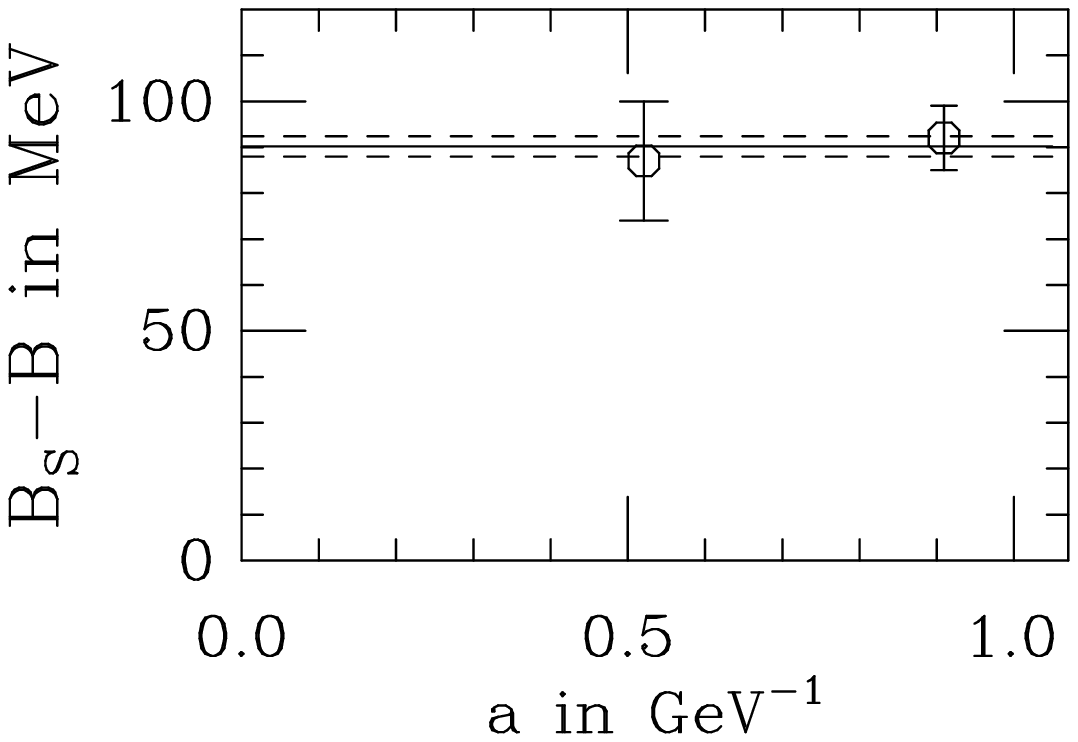,
width=3.7cm,bbllx=35pt,bburx=350pt,bblly=55pt,bbury=265pt}}
{\protect\epsfig{file=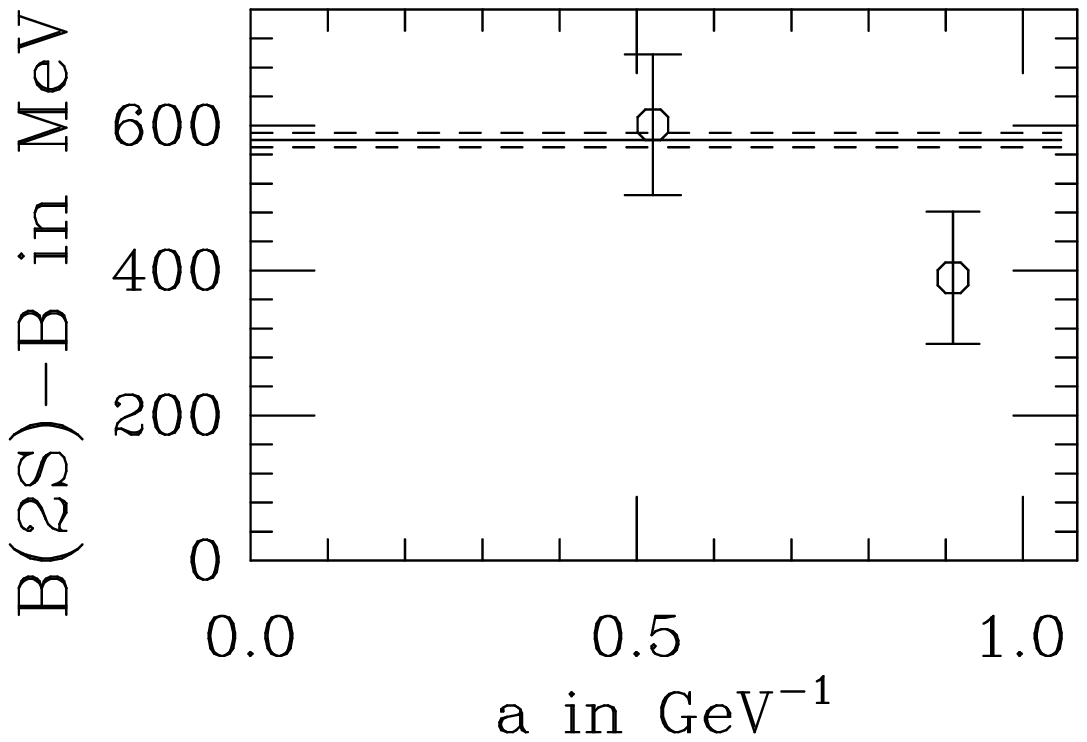,
width=3.7cm,bbllx=35pt,bburx=350pt,bblly=55pt,bbury=265pt}}}
\protect\caption{\label{spec_scaling} Scaling of $B$-meson splittings.
Please note the different nature of these quantities. From the left we
give an example for a spin dependent, a flavour dependent and a radial
splitting. Experimental results apart from the $B(2S)$ are
from the PDG~\protect\cite{pdg}. The $B(2S)$ is a recent result 
from DELPHI~\protect\cite{delphi}.}
\end{figure}

Figure~\ref{spec_scaling} shows results obtained at two
different values of the lattice spacing~\cite{arifaspec,beta57} in the
quenched theory. 
No sign of large lattice spacing dependence can be observed for any
of the quantities. 
For the spin independent results one observes reasonable agreement with
the experimental result, however the results for the $B_s$-hyperfine is
in clear disagreement. Here the discussion of section~\ref{upsplit}
also applies. The quenched approximation seems to have an effect as well, 
since the heavy clover approach with different systematic errors,
delivers a similar reduction of the hyperfine~\cite{boyle}. 

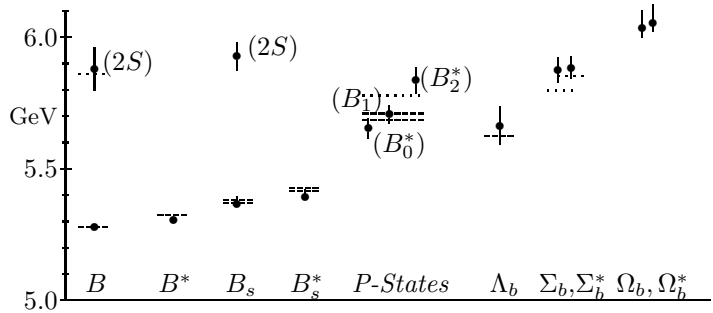
\begin{figure}[t]
\centerline{\input{arifaspec}}
\protect\caption{\label{hlspecfig} Spectrum of mesons and baryons containing
one heavy $b$-quark in the quenched approximation
\protect\cite{arifaspec}. Dashed lines indicate experimental results
from the PDG~\protect\cite{pdg} and dotted lines recent
results from DELPHI~\protect\cite{delphi}. }
\end{figure}

Figure \ref{hlspecfig} gives an overview over the hadron spectrum
containing one heavy $b$-quark.
Apart from the above discussed hyperfine splittings one observes 
good agreement with experimental results, including heavy
baryons. 

\subsection{Pseudoscalar decay constant}
Because of the small leptonic branching fraction, the pseudoscalar
decay constant $f_B$ is hard to measure experimentally, however its
knowledge is required in the determination of e.g.\ the
bag parameter in $B$-$\bar B$ mixing.

\begin{figure}[tp]
\centerline{\protect\epsfig{file=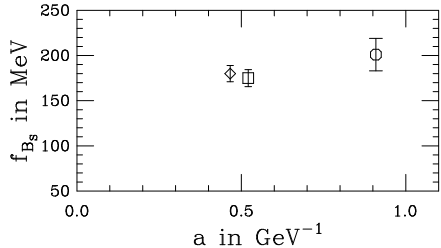,width=6.8cm,
bbllx=7pt,bburx=570pt,bblly=46pt,bbury=325pt}}
\protect\caption{\label{fbscale} Scaling of $f_{B_s}$. The
square~\protect\cite{arifafb} and octagon~\protect\cite{fbsnew} use 
higher order corrections in the Hamilton than the
diamond~\protect\cite{ishikawa}. The latter
has been scaled to match the $a$-value determined from $m_\rho$
 used in the other ones. Errors
encompass the statistical uncertainties and those arising out of the
operator renormalisation and mixing~\protect\cite{colinjunko}. 
In case of the diamond an estimate of the effect of the lower order 
Hamiltonian is also included.}
\end{figure}

In order to obtain reliable results the inclusion of
operator renormalisation and mixing is 
crucial~\cite{colinjunko,beta57,arifafb,ishikawa}.
In figure~\ref{fbscale} we compile a scaling plot for $f_{B_s}$.
The plot shows good scaling as the lattice spacing is changed. 
We quote the result of reference~\cite{arifafb} which is in good
agreement with the findings of reference~\cite{ishikawa}:
\bgeq
f_B=147(11)(^{+8}_{-12})(9)(6){\rm\ MeV}\,,\fin
f_{B_s}=175(8)(^{+7}_{-10})(11)(7)(^{+7}_{-0}){\rm\ MeV}\,.
\edeq
These results also agree with those of other lattice calculations
using Dirac type Hamiltonians~\cite{draper}.

\section{Conclusion and outlook}
We discussed phenomenological heavy quark results including
spectroscopy and  the decay constants $f_B$ and
$f_{B_s}$. Future work has to improve on spin dependent
splittings and quenching effects have to be further addressed.

\section*{Acknowledgements}
It is a pleasure to thank Christine Davies, Arifa Ali Khan, Sara
Collins and Achim Spitz for suggestions and support while preparing
this talk. A Marie Curie research fellowship 
by the European commission under ERB FMB ICT 961729
is gratefully acknowledged.

\end{document}

%% file: arifaspec.tex
 
\setlength{\unitlength}{0.035cm}
\begin{picture}(260,110)(0,500)


\put(15,500){\line(0,1){110}}
\multiput(13,500)(0,50){3}{\line(1,0){4}}
\multiput(14,500)(0,10){12}{\line(1,0){2}}
\put(12,500){\makebox(0,0)[r]{{5.0}}}
\put(12,550){\makebox(0,0)[r]{{5.5}}}
\put(12,600){\makebox(0,0)[r]{{6.0}}}
\put(12,570){\makebox(0,0)[r]{{\small GeV}}}
\put(15,500){\line(1,0){245}}


     \put(25,510){\makebox(0,0)[t]{{ $B$}}}
     \put(26,528){\circle*{3}}
     \multiput(20,527.9)(3,0){4}{\line(1,0){2}}
     \put(26,587.9){\circle*{3}}
     \put(26,589.4){\line(0,1){6.8}}
     \put(26,586.4){\line(0,-1){6.8}}
     \put(37,595){\makebox(0,0)[t]{{ $(2S)$}}}
     \multiput(20,586)(3,0){4}{\line(1,0){0.5}}

     \put(55,510){\makebox(0,0)[t]{{ $B^{*}$}}}
     \put(56,530.4){\circle*{3}}
     \put(56,530.4){\line(0,1){0.5}}
     \put(56,530.4){\line(0,-1){0.5}}
     \multiput(50,532.6)(3,0){4}{\line(1,0){2}}
     \multiput(50,532.4)(3,0){4}{\line(1,0){2}}

     \put(80,510){\makebox(0,0)[t]{{ $B_s$}}}
     \put(80,536.7){\circle*{3}}
     \put(80,536.7){\line(0,1){2.9}}
     \put(80,536.7){\line(0,-1){1}}
     \multiput(75,538.1)(3,0){4}{\line(1,0){2}}
     \multiput(75,536.9)(3,0){4}{\line(1,0){2}}
     \put(80,592.8){\circle*{3}}
     \put(80,594.3){\line(0,1){3.8}}
     \put(80,591.3){\line(0,-1){3.8}}
     \put(92,601){\makebox(0,0)[t]{{ $(2S)$}}}

     \put(105,510){\makebox(0,0)[t]{{ $B^{*}_s$}}}
     \put(106,539.4){\circle*{3}}
     \put(106,539.4){\line(0,1){2.9}}
     \put(106,539.4){\line(0,-1){1.0}}
     \multiput(100,542.8)(3,0){4}{\line(1,0){2}}
     \multiput(100,541.6)(3,0){4}{\line(1,0){2}}

     \put(142,510){\makebox(0,0)[t]{{\it P-States}}}
     \put(148,583.7){\circle*{3}}
     \put(148,585.2){\line(0,1){3.5}}
     \put(148,582.2){\line(0,-1){3.5}}
     \put(158,584.2){\makebox(0,0){{ $(B^*_2)$}}}

     \put(130,565.3){\circle*{3}}
     \put(130,566.8){\line(0,1){2.3}}
     \put(130,563.8){\line(0,-1){2.3}}
     \put(140,564.0){\makebox(0,0)[t]{{ $(B^*_0)$}}}
     \put(138,570.7){\circle*{3}}
     \put(138,572.2){\line(0,1){1.9}}
     \put(138,569.2){\line(0,-1){1.9}}
     \put(124,576){\makebox(0,0){{ $({B_1})$}}}
     \multiput(128,568.6)(3,0){8}{\line(1,0){2}}
     \multiput(128,571.0)(3,0){8}{\line(1,0){2}}
     \multiput(128,577.9)(3,0){8}{\line(1,0){0.5}}

     \put(180,510){\makebox(0,0)[t]{{ $\Lambda_b$}}}
     \put(180,566.5){\circle*{3}}
     \put(180,566.5){\line(0,-1){7.1}}
     \put(180,566.5){\line(0,1){7.1}}
     \multiput(174,562.4)(3,0){4}{\line(1,0){2}}

     \put(206,510){\makebox(0,0)[t]{ {$\Sigma_b$},$\Sigma_b^*$}}
     \put(202,587.6){\circle*{3}}
     \put(202,587.6){\line(0,1){4.6}}
     \put(202,587.6){\line(0,-1){4.6}}
     \multiput(198,579.7)(3,0){4}{\line(1,0){0.5}}

     \put(207,588.5){\circle*{3}}
     \put(207,588.5){\line(0,1){4.3}}
     \put(207,588.5){\line(0,-1){4.3}}
     \multiput(202,585.3)(3,0){4}{\line(1,0){0.5}}

%
     \put(238,510){\makebox(0,0)[t]{{$\Omega_b,\Omega_b^{*}$}}}
     \put(234,603.4){\circle*{3}}
     \put(234,603.4){\line(0,-1){3.3}}
     \put(234,603.4){\line(0,1){6.8}}
     \put(238,605.4){\circle*{3}}
     \put(238,605.4){\line(0,-1){3.3}}
     \put(238,605.4){\line(0,1){7.1}}
\end{picture}